\documentclass[12pt]{article}

\newcommand{\beq}{\begin{equation}}
\newcommand{\eeq}{\end{equation}}
\newcommand{\bea}{\begin{eqnarray}}
\newcommand{\eea}{\end{eqnarray}}
\newcommand{\du}{d_{\cal U}}

\newcommand{\lagr}{{\cal L}}
\newcommand{\opr}{{\cal O}}
\newcommand{\unp}{{\cal U}}
\newcommand{\bra}{\langle}
\newcommand{\ket}{\rangle}

\usepackage{graphicx}
 \usepackage{setspace}  
\begin{document}
\vskip 3cm
\begin{center}
{\bf \LARGE Unparticle-Enhanced  Black Holes at the LHC}

\vskip .4cm
J. R. Mureika \\
{\small \it Department of Physics, Loyola Marymount University, Los Angeles, CA  90045-2659\\
Email: jmureika@lmu.edu}
\end{center}
\vskip 4cm

{\noindent{\bf Abstract} \\
Based on the idea that tensor unparticles can  enhance the
gravitational interactions between standard model particles, potential black hole formation in high energy collisions is examined.  Modifications to the horizon radius $r_H$ are derived, and the corresponding geometric cross-sections of such objects are calculated.  It is shown that $r_H$ increases dramatically to the electroweak scale for masses $M_{BH} \sim 1-10~$TeV, yielding a geometric cross-section on the order of $\leq$50~pb.   This suggests that unparticle physics provides a mechanism for black hole formation in future accelerators, without the requirement of extra spatial dimensions. }

\pagebreak

\section{Introduction}
Anticipation of the stream of collider data to flow from the LHC has seen theorists uncovering a plethora of new physics that might be lurking just beyond the TeV-energy scale.    This had led to the prediction of new extensions to particle physics, as well as modifications to accepted general physical laws.   One of the most popular and innovative of these proposals is the possible existence of large sub-millimetre extra dimensions that could radically alter phenomenology at the electroweak scale \cite{add}.

Recently, it was proposed that there could be a scale-invariant high-energy particle sector of unknown composition with a non-trivial fixed point \cite{georgi}.  This would normally be weakly-coupled to the standard model, but below some threshold energy scale -- possibly in the TeV range -- a dimensional transmutation in the hidden sector operator would allow for stronger interactions.  Dubbed ``unparticle physics'' because of the non-intuitive phase space structure, its introduction has once again caused a flurry of research into modifications to known physics, including high energy particle phenomenology \cite{unphep}, astrophysical and cosmological events \cite{lewis, mcdonald, unpast}, and low to ultra-high energy neutrino physics \cite{unpnu}.  Basically, unparticle physics stands to modify high energy events in the same spirit of large extra dimensions, and signatures of unparticle-driven events at the LHC have largely been the topic of choice in the literature.
 
One of the most intriguing effects from large extra dimensions is the possibility of black hole production at the LHC and other future accelerators, originally proposed by the authors of \cite{fischler} and later expanded upon (see References \cite{bh1,bh3,kanti} and others cited witin).  The unusually large gravitational attraction at scales below the dimensional compactification scale will inflate the size of the horizon distance of TeV-sized masses, allowing for the potential of copious numbers of such events.  These black holes will evaporate quickly, but will be observable primarily through their unique Hawking temperature signature and subsequent decay shower.  Unparticle-driven interactions will also alter the classical laws of gravitation, and several papers have examined the possible limits of observational signatures that might result in table-top Cavendish-like experiments \cite{goldberg}, as well as long range \cite{deshpande} and solar system-scale effects in orbital precession \cite{damora}.   

To date, there has been no attempt to estimate the likelihood that the enhanced gravitational attraction from unparticle interactions can play a role in black hole formation, particularly at the TeV-scale.  This note will thus examine such a possibility, subject to the known properties of tensor unparticles.  A review of the unparticle foundations and literature will be given, followed by a discussion of higher-dimensional black holes, and their creation in particle collisions.  The unparticle black hole formalism will then be outlined, and the likelihood of black hole formation in the LHC will be discussed.

\section{An overview of unparticle physics}
At the time this manuscript was written, there were approximately 90 papers on unparticle physics theory and phenomenology.  A full citation of the existing literature would be exhaustive, so only key references to related subjects are provided herein.  The interested reader is urged to follow up on the appropriate paper, or search the appropriate on-line databases. 

The inspiration for unparticles physics, as outlined in the pioneering papers by Georgi \cite{georgi}, is derived from the Banks-Zaks (BZ) high energy field theory with non-trivial IR fixed point, whose interactions are mediated by exchange particles of mass $M_{\unp}$ \cite{bz1}.  Such a field is assumed to interact with the Standard Model (SM) fields according to the scale-suppressed non-renormalizable Lagrangian density
\beq
\lagr = \frac{1}{M^k_{\unp}} \opr_{SM}\opr_{BZ}
\eeq
where the $\opr$ represent the respective field operators of dimensions $d_{SM}$ and $d_{BZ}$ respectively, and $k = d_{SM} + d_{BZ} - 4$ to ensure a dimensionless action.  Below some other energy scale $\Lambda_\unp < M_\unp$, the Banks-Zaks operator undergoes dimensional transmutation to a new ``unparticle'' operator $\opr_{BZ} \rightarrow C_\unp \lambda^{d_{BZ} - d_\unp}\opr_\unp$ of dimension $d_\unp \ne d_{BZ}$, yielding a structurally-identical Lagrangian
\beq
\lagr = \frac{\kappa}{\Lambda^{k_\unp}} \opr_{SM} \opr_\unp~~~,~~~\kappa = C_\unp \left(\frac{\Lambda_\unp}{M_{BZ}}\right)^k
\eeq
with $k_\unp = d_{SM}+d_\unp-4$.

From the propagator matrix element
\beq
\bra 0 | \opr_\unp(x) \opr^\dagger_\unp(0)|0\ket = \int \frac{d^4p}{(4\pi)^4} e^{iPx} |\bra 0 | \opr_\unp(0)|P\ket|^2 \rho(P^2)
\eeq
for an unparticle of four-momentum $P$, the spectral density function can be shown to have the form
\beq
 |\bra 0 | \opr_\unp(0)|P\ket|^2 \rho(P^2) = {A_d}_{\cal U} \theta(P^0) \theta(P^2) (P^2)^{\du -2}
\label{unpphase}
\eeq
with
\beq
{A_d}_{\cal U} = \frac{16 \pi^{5/2}}{(2\pi)^{2n}} \frac{\Gamma(n+1/2)}{\Gamma(n-1)\Gamma(2n)}~~.
\eeq
Equation~\ref{unpphase} is identical in form to the  analogous phase space derived from the interactions of $n$ particles of total momentum $P$,
\beq
A_n \theta(P^0) \theta(P^2) (P^2)^{n -2}~~,
\eeq
which leads to the interpretation that unparticles ``resemble'' a collection of non-integer ($\du$) indivisible particles.  It should be noted that this conclusion is meant only to exemplify the non-intuitive nature of unparticle matter, much in the same spirit as describing a physical fractal distribution by a non-integer scaling dimension.  Just as in reality there cannot be a non-integer number of spatial dimensions, it is highly unlikely that unparticles actually come in fractional numbers.

Since we have yet to observe unparticle signatures in accelerator experiments, the transmutation scale must be at least $\Lambda_\unp > 100~$GeV.  Of course, this leaves door open to $\Lambda_\unp  \sim 1~$TeV, in which case the LHC and other future collider experiments may well represent the testbed for such a  theory.  The free parameters $\du$, $d_{BZ}$ and $M_\unp$ can then be used to constrain the unparticle interactions to experimental and observational (cosmological) data.  

The standard limits in the literature on the unparticle dimension are $\du > 1$, with the upper bound depending on the form of the operator $\opr_\unp$.  In Reference \cite{mcdonald}, it is shown that $M_\unp \geq 20-2600~$TeV for $1.1\leq \du \leq 2$, based on cosmological considerations.    The author of \cite{lewis} places more liberal limits on the mass of the echange particle from supernovae cooling rates due to electron-positron and photon-photon annihilation to tensor unparticles, suggesting that $M_\unp \geq 1.8-180~$TeV for $4/3 \leq \du \leq 2$.  From the gravitiational perspective, it is shown \cite{goldberg} that deviations to the inverse-square law will manifest themselves on the sub-millimetre scale if $M \sim 5-50~$TeV for $\du = 2$, but with $M_\unp$ growing significantly large for smaller $\du$.  The value of $d_{BZ}$ is unknown, but it is generally accepted that it should fall around unity \cite{damora}.  Reference \cite{mcdonald} shows how variation of $d_{BZ}$ between 1 and 3, showing that as $d_{BZ}$ increases, lower bounds on $M_\unp$ are possible, albeit with larger sensitivity to $\Lambda_\mu$.

As the framework of unparticle physics is an effective field theory, the exact form of the operators $\opr_\unp$ are unknown.  It is assumed that their algebraic properties and associated Lorentz structures mimic those of conventional particles, though, and thus one can expect scalar, vector, tensor, and spin unparticle interactions.  Of particular relevance to gravity are tensor unparticle interactions that can couple to (and re-scale) the stress-energy tensor by a perturbation
\beq
T^{\mu\nu} + T_\unp^{\mu \nu}~~,~~ T_\unp^{\mu \nu} \sim  \sqrt{|g|}T^{\alpha \beta}\opr^\unp_{\alpha \beta} \; g_{\mu \nu}
\eeq
introducing a cosmological constant-like term to the action \cite{goldberg}.  This coupling implies that $\opr^\unp_{\mu \nu}$ must represent a spin-2 particle, which one could dub an ``ungraviton''.  Vector unparticle operators $\opr_\unp^\mu$ may couple to baryon currents $B_\mu$ via interactions of the form $B_\mu \opr_\unp^\mu$, yielding a repulsive couping that serves to decrease the effective strength Newtonian gravity \cite{deshpande}.

In the non-relativistic limit, the appropriate interactions may be computed by the usual Fourier transform method.  For tensor couplings, the result is a modified potential of the form \cite{goldberg,damora}
\beq
V(r) = V_N(r) \left[1 + \frac{2}{\pi^{2\du-1}}\; \frac{\Gamma(\du + \frac{1}{2})\Gamma(\du -\frac{1}{2})}{\Gamma(2\du)} \left(\frac{R_*}{r}\right)^{2\du-2}\right] = V_N(r) \left[1+\Gamma_{\du}\left(\frac{R_*}{r}\right)^{2\du-2}\right] ~ ,
\label{tensorV}
\eeq
where $V_N(r)$ is the usual Newtonian potential.  The effective length-scale $R_*$ is defined as 
\beq
R_* = \Lambda_\unp^{-1} \left(\frac{M_{Pl}}{\Lambda_\unp}\right)^\frac{1}{\du-1} \left(\frac{\Lambda_\unp}{M_\unp}\right)^\frac{d_{BZ}}{\du-1}~~.
\eeq 

If $\du < 1$ solutions are allowed, then the unparticle potential can be repulsive, since $\Gamma(x) < 1$ for $x <0$ (and hence for $\du < 1/2$).  In general, most unparticle treatments in the literature do not address the latter case.  The tensor rank of the operator ${\cal O}_\unp$ will determine the dimensionality $\du$.   

Following the completion of this paper, it was found that unparticle interactions with normal matter introduce much more restrictive constraints on the possible values of the dimension $\du$ \cite{grinstein}.  For vector interactions, unitarity is violated unless $\du \geq 3$, while for tensor unparticles $\du \geq 4$ is required.   It was also suggested that the forms of the propagators for vector and tensor unparticles in the existing literature should be modified, with the interactions reflecting new contact terms between standard model particles in addition to the unparticle exchange.   For vector unparticles, this has the effect of re-weighting terms in the propagator as a function of unparticle dimension $\du$, and can yield corrections to the pure unparticle effects.  Similar effects were noted for tensor unparticle interactions.

The results in this paper reflect the unitarity constraints imposed in Reference~\cite{grinstein}, using the propagator introduced in \cite{goldberg}.  As such, these results should be considered  a first-order estimate of unparticle gravitational effects at the TeV scale.  Accounting for any changes to the propagator is left to future works.

\section{Black holes in (4+n)-dimensional space}
The usual Schwarzschild radius $r_S = 2m/M_{Pl}^2$ for most objects is exceedingly small, due primarily to the enormity  of the Planck scale.  It has been suggested that this hierarchy problem may be solved through the introduction of large extra dimensions \cite{add} into which only gravity  may propagate, and that the actual Planck scale may not be that dissimilar from the known electroweak energies of order $\sim 1~$TeV.  One of the many interesting phenomenological consequences of extra dimensional theories with low Planck scale is the formation of black holes.  The ``classical'' value of $r_S$ is strikingly altered by their presence, due to the graviton leakage into the bulk.  The Schwarzschild metric in $4+n$-dimensional spacetime is \cite{kanti}
\beq
ds^2 = \left(1-h(r)^{n+1} \right)\; dt^2 - \frac{dr^2}{1-h(r)^{n+1}} - r^{2} d\Omega^2_{n-2}
\label{addbh}
\eeq
with
\beq
h(r)  = \frac{r_H}{r}~~.
\eeq
In the semi-classical limit, the modified horizon distance $r_H$ can be shown to be \cite{bh1}
\beq
r_H = \frac{1}{\sqrt{\pi}M_{*}} \left(\frac{M_{BH}}{M_{*}}\right)^{\frac{1}{n+1}} \left(\frac{8\Gamma\left(\frac{n+3}{2}\right)}{n+2}\right)^{\frac{1}{n+1}}
\label{addrh}
\eeq
for a black hole of mass $M_{BH}$ subject to a gravitational scale $M_*$.  In extra dimensional models where $M_* \sim 1~$TeV ({\it i.e.} $M_* \ll M_{Pl}$), the resulting black holes are much larger than their four-dimensional spacetime counterparts.  These differences become quite pronounced when the mass is small, and close to (but still greater than) the new Planck scale, $M_{BH} > M_*$.   This furthermore implies a distinct signature for the black hole's Hawking radiation spectrum and decay modes, which can likely be detected at the LHC.

A mini-black hole will be formed if two colliding partons with energy $\sqrt{s} > 1~$TeV pass within an impact parameter $b \leq r_H$, or alternatively interact within the geometric cross-section $\sigma_{BH} \sim \pi r_H^2$.    The latter can serve as a rough approximation for the production cross-section in the parton collisions.  For a mass $M_{BH} \sim 1~$TeV, the standard Schwarzschild radius is $r_H = r_s \sim 10^{-31}~$TeV$^{-1}$, or about $10^{-50}~$m.  The chance of such black hole formation in the ``standard'' theory is thus zero.   In the case of ADD large extra dimensions, it has been shown that the horizon radius for TeV-scale events may be increased to about $10^{-4}~$fm, or a cross-section of $\sigma \sim 100~$pb (see \cite{fischler,bh1,bh3,kanti} and references therein).  Although the structure of black holes becomes ``stringy'' and vastly non-trivial as $M_{BM} \sim M_{Pl}$, the general consensus in the literature is that the classical approximation of the horizon distance  is adequate for discussion.

\section{Tensor unparticle-enhanced black hole formation}
\label{unpbh}
From the definitions of the modified gravitational potential energies in Equations~\ref{tensorV} and ~\ref{vecV}, it is presupposed that one may re-define the gravitational {\it potential} of a single mass $m$, according to the standard definition.  The modified gravitational potentials for both tensor and vector couplings can be written in the general form
\beq
\Phi_\unp(r) = \Phi_N(r)\left[1+\Gamma_{\du} \left(\frac{R_*}{r}\right)^{2\du-2}\right]~,
\eeq
The following analysis will consider modifications to the Schwarzschild metric due to unparticle interactions that mimic enhancements to the gravitational potential.   In that sense, it is instructive to re-write the $\mu=\nu = 0,1$ Schwarzschild metric terms as \cite{weinberg}
\beq
g_{00} = 1+2\Phi(r)~~~;~~~g_{11} = \left(1+2\Phi(r)\right)^{-1}
\label{schpot}
\eeq
This is an appropriate expression for any number of spacetime dimensions, and can be understood to incorporate unparticle interactions for two reasons.   First, the modified Newtonian potentials should be recovered in the weak-field limit.  Secondly,  for interactions near the scale $R_*$, the unparticle contribution will dominate and the metric will resemble that of a ``$(2\du-2)$-dimensional'' Schwarzschild solution. Thus, the metric can now be written in the form
\bea
ds^2 = \left[1-\frac{2GM}{r}\left(1+ \Gamma_{\du} \left(\frac{R_*}{r}\right)^{2\du-2}\right)\right] \; dt^2 + 
 \frac{dr^2}{1-\frac{2GM}{r}\left(1+\Gamma_{\du} \left(\frac{R_*}{r}\right)^{2\du-2}\right)} + r^2 d\Omega^2 ~~.
\label{unpmetric}
\eea
 A full derivation of the unparticle-enhanced $g_{\mu\nu}$ stemming from unparticle corrections to the stress-energy tensor would ultimately shed definitive light on the situation.

The value of the mass scale $M_\unp$ and dimension $\du$, along with the BZ operator dimension $d_{BZ}$, will uniquely determine $R_*$.  The value of $d_{BZ}$ is most likely not significantly different from unity, so for the purposes of simplicity one can assume $d_{BZ} \sim 1$.  Setting $\Lambda_\unp = 1~$TeV, the scale length simplifies to
\beq
R_* = \left(\frac{M_{Pl}}{M_\unp}\right)^\frac{1}{\du-1}~{\rm TeV}^{-1} 
\label{unpscale}
\eeq
This relation is algebraically similar to the order-of-magnitude estimate for the size of extra dimensions \cite{add}, with $M_\unp$ taking the place of $M_*$.  Since it is likely that $M_\unp \ll M_{Pl}$, this result implies $R_*$ will be unusually large and will ultimately dominate the solutions for the ungravity-modified horizons $r_H$ if $r_s \ll R_*$.

Of particular interest in this regard will be the formation of mini-black holes in future accelerator experiments.  For the purposes of this analysis, a similar assumption for the formation of a semi-classical black hole in the extra-dimensional case -- namely $M_{\rm BH} > M_{*}$ --  will hold with ungravity-ehanced singularity creation.  The ``threshold'' value of $M_{BH}$ can be obtained by approximating the potential as the dominant unparticle contribution
\beq
\Phi(r) \sim \frac{GM_{BH}\Gamma_{\du}}{r} \left(\frac{R_*}{r}\right)^{2\du-2}~~.
\eeq
Inserting this into the metric and solving for the horizon radius gives an $M_{Pl}$-free expression
\beq
r_H \approx \left(\frac{2M_{BH}\Gamma_{\du}}{M_\unp^2 \Lambda_\unp^{-1}}\right)^\frac{1}{2\du-1}\; \Lambda_\unp^{-1}
\label{approxcs}
\eeq
In this case, the ungravity-enhanced interactions will yield black hole formation of mass $M_{\rm BH}  =M_\unp^2\Lambda_\unp^{-1}$.   For $\Lambda_\unp \sim 1~$TeV, $M_\unp \sim 10~$TeV, and $M_{BH} \sim 10~$TeV, one finds $r_H \sim 10^{-5}~$fm for all values of $\du \geq 4$, which corresponds to a geometric cross section $\sigma_{BH} \sim 10~$pb.   This places the likelihood of unparticle-driven black hole formation at the LHC in a favorable light.

\section{Results and Discussion}
\begin{figure}[h]
\begin{center}
\leavevmode
\includegraphics[scale=0.6]{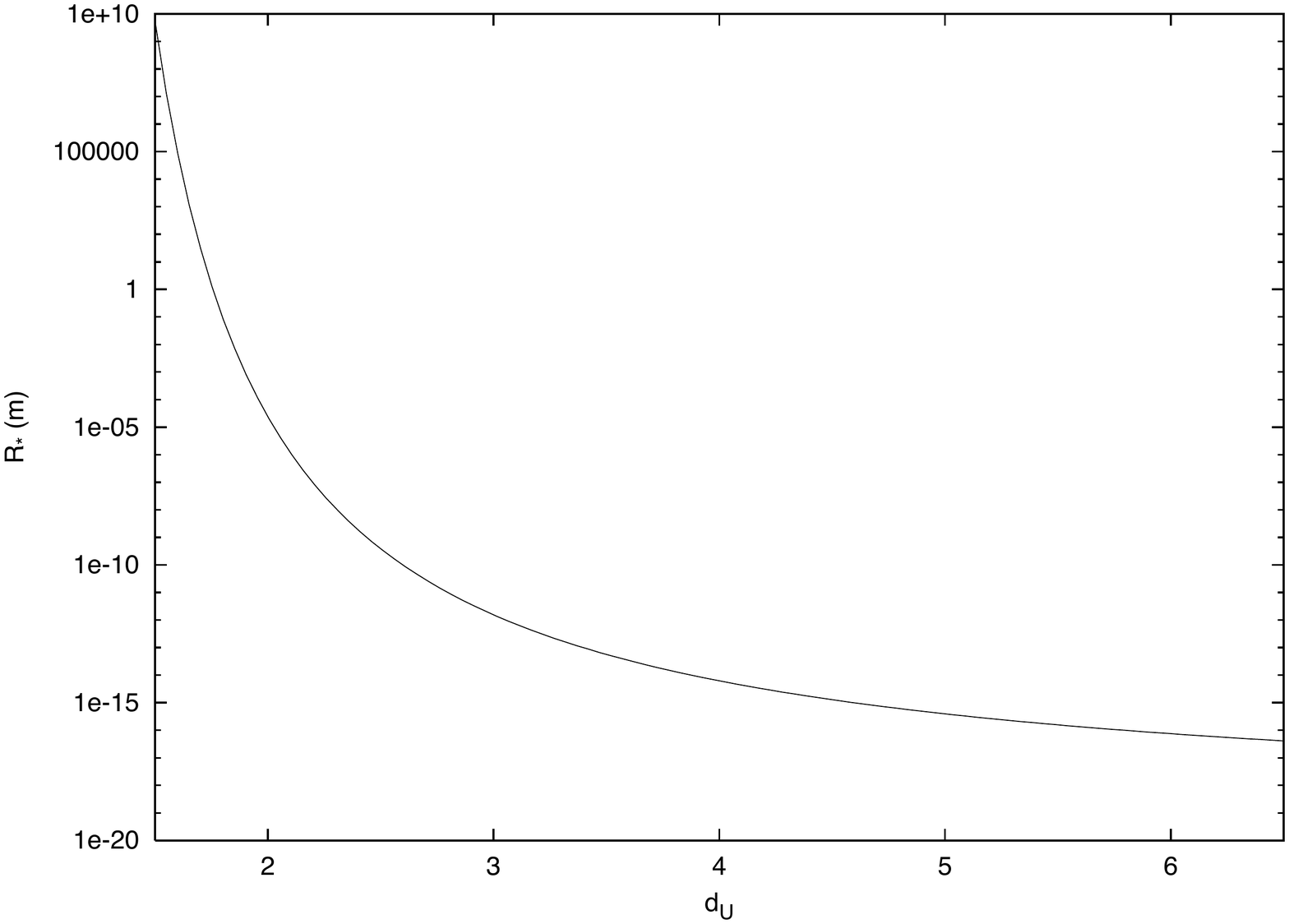}
\caption{Unparticle interaction scale $R_*$ (in metres) as a function of dimension $\du$, with $\Lambda_\unp = 1~$TeV, $M_\unp = 10~$TeV, and $M_{Pl} = 2.4\times 10^{15}~$TeV.}
\label{fig1}
\end{center}
\end{figure}

Numerically-speaking, the goal is thus to see if the mechanism can increase the standard horizon radius by roughly 30 orders of magnitude, to be commensurate with accelerator scales.   At the same time, the value of the interaction scale $R_*$ should be below constraints imposed by current experiments.   As Figure~\ref{fig1} demonstrates, values of $\du \geq 2$ produce such conditions, with $\Lambda_\unp \sim 1~$TeV and $M_\unp \sim 10~$TeV. 

An unparticle sector of dimension $\du=1$ will induce a correction to the usual Schwarzshild solution of the form
\beq
r_H = r_s(1+\Gamma_1)~.
\eeq
Since $\Gamma_1 = 1$, this will always result in a modification of the ``standard'' solution by a factor of 2, and thus can be absorbed into a re-scaled gravitational constant.   This solution is thus not of particular interest.  Furthermore, this is excluded by unitarity constraints \cite{grinstein}.

\begin{figure}[h]
\begin{center}
\leavevmode
\includegraphics[scale=0.6]{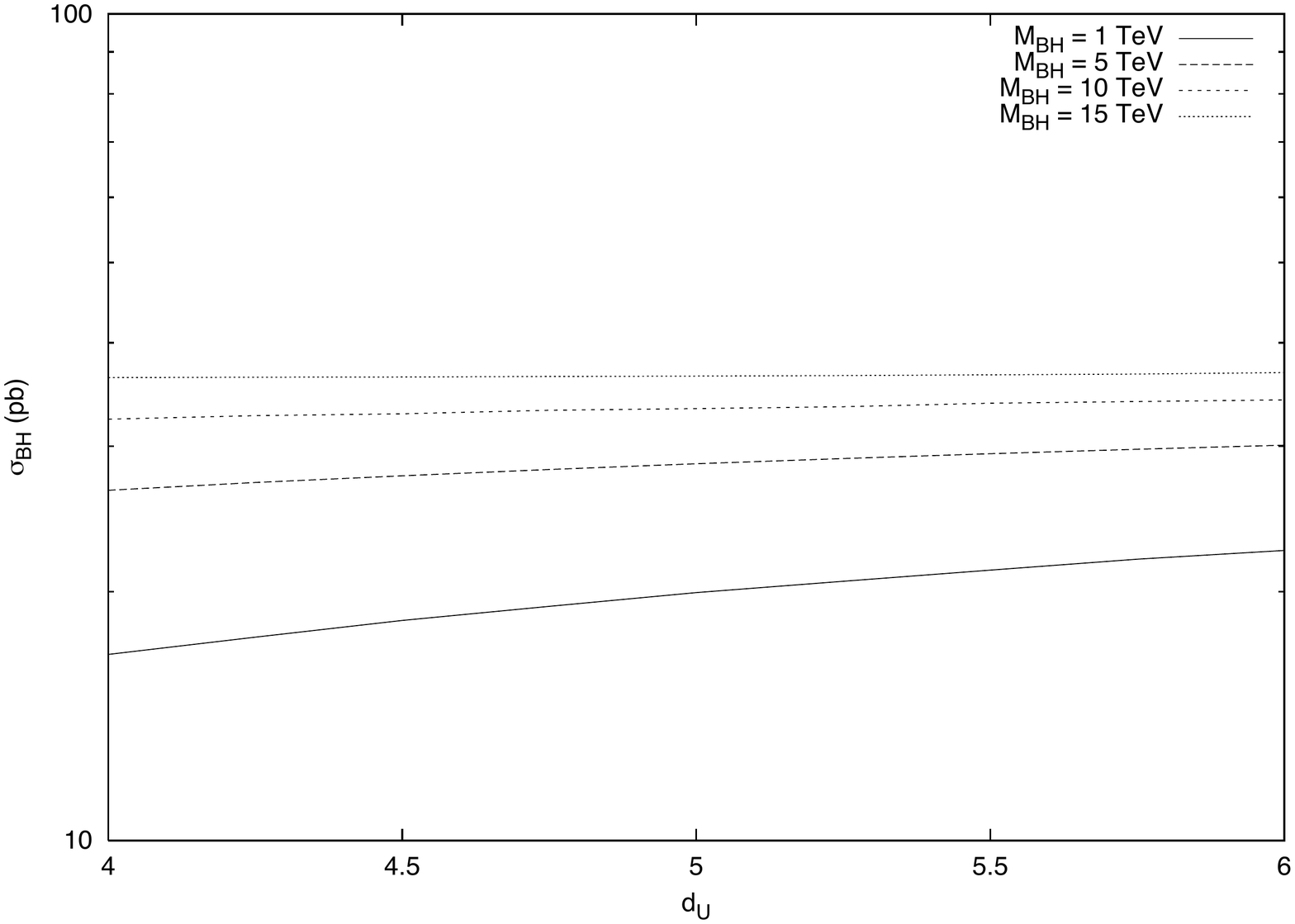}
\caption{Geometric black hole cross-section $\sigma_{BH} = \pi r_H^2$ as a function of dimension $\du$, with $\Lambda_\unp = 1~$TeV, $M_\unp = 10~$TeV, and $M_{Pl} = 2.4\times 10^{15}~$TeV.}
\label{fig2}
\end{center}
\end{figure}

As the unparticle dimension grows, however, the factor $(M_{Pl}/M_\unp)^\frac{1}{\du-1}$ is re-introduced and the solutions become greatly affected.   The horizon condition will in general yield multiple solutions (real and complex), but only the positive values are taken to have physical significance.  The horizon radii $r_{H}$ for various values of $\du$ and $M_{BH}$ have been solved using the singularity condition imposed on Equation~\ref{unpmetric}, and are displayed in Figure~\ref{fig2}.   The desired dimensional range for a tensor unparticle operator coupling to the gravitational sector is  $\du \geq 4$.

The prognosis for unparticle-enhanced black hole creation at the LHC or other future TeV-scale collider is hopeful.  The most promising area of parameter space that would allow for black hole formation is when $\du$ is large and $M_\unp \sim 10~$TeV.   For this range of values, the geometric cross-section $\sigma_{BH}$ is in the range $\sigma \sim 10-100~$pb, with higher values being caused by larger values of $M_{BH}$.    Note also that for $\du\geq2$ (relevant to tensor unparticles), the interaction scale is in the range $R_* < 10^{-12}~$m, which is much smaller than the current limits on deviations from Newtonian gravity. 

The largest cross-sections ($\sigma_{BH} \sim 40~$pb) arise for $M_{BH} = 15~$TeV, corresponding to the maximum possible center-of-mass energy at the LHC.  Since the actual collisions will have $\sqrt{s}$ much less than this, it is reasonable to assume that the possible values of $\sigma_{BH}$ are constrained to lie in the parameter space between the upper and lower lines of Figure~\ref{fig2}, or $\sigma_{BH} \sim 10~$pb, in agreement with the approximation derived from Equation~\ref{approxcs}.

Due to the form of $R_*$ given in Equation~\ref{unpscale}, it can been shown that as the scale $M_\unp$ grows and $R_*$ approaches the electroweak regime.  The corresponding cross-sections would thus suggest a very low probability of event observation.  The values in Figure~\ref{fig2} can be taken to be an upper limit on the mechanism parameters, since increasing $M_\unp$ and $\Lambda_\unp$ will result in vastly smaller cross-sections.

Experimentally, if such black holes are created at the LHC they will occur with lower frequency than for their extra-dimensional counterparts.  The latter has been liberally estimated as a 1~Hz production for a beam luminosity of 30~fb$^{-1}/$y and cross section of 100~pb \cite{bh3}.  Using the $\sigma_{BH} < 50$~pb figures culled from this analysis as an ideal interaction cross-section, unparticle-enhanced black holes should be produced with a similar frequency, at least to within an order of magnitude (hence a black hole every few seconds).  These holes are also potentially distinguishable via their Hawking temperature, which scales as the inverse horizon radius:
\beq
T_H = \frac{1}{4\pi r_H} = \frac{\Lambda_\unp}{4\pi} \left(\frac{2M_{BH}\Gamma_{\du}}{M_\unp^2 \Lambda_\unp^{-1}}\right)^{-\frac{1}{2\du-1}}
\eeq
The possible temparture spectra are much richer than those for extra-dimensional black holes, which grow as $T_H \sim M_* \left(M_{BH}/M_*\right)^{-1/(1+n)}$, with $n$ restricted to integer values.

\section{A comment on vector unparticle enhanced black holes}

If the $\opr_\unp$ is vector-like, the unparticles couple to baryon currents with (dimensionless) strength $\lambda_B$ according to the interaction
\beq
\lagr= \lambda_B \Lambda_\unp^{1-\du} B_\mu \opr^\mu_\unp~
\eeq
will yield a new interaction potential of the form \cite{oregon}
\beq
V_\unp(r) \sim \frac{\lambda_B B_1 B_2}{r^{2\du-1}} \longrightarrow \frac{\lambda_B m_1 m_2}{u^2 r^{2\du-1}}
\eeq
where the brayon numbers for the interacting masses are $B_j \approx m_j/u$.  The modified gravitational potential is then
\beq
\Phi(r) = \Phi_N(r) \left[ 1 - \frac{1}{2\pi^{2\du}}\frac{\Gamma(\du+\frac{1}{2})\Gamma(\du-\frac{1}{2})}{\Gamma(2\du)} \left(\frac{R_{*v}}{r}\right)^{2\du-2} \right] = \Phi_N(r) \left[ 1- \bar{\Gamma_d} \left(\frac{R_{*v}}{r}\right)^{2\du-2}\right]
\label{vecV}
\eeq
with the new length scale $R_{*v}$ dependent on the coupling strength $\lambda_B$ and the other unparticle parameters.

Since vector unparticle couplings to baryons will result in a repulsive correction to gravity, such a mechanism will not contribute to black hole formation if $\du \geq 1$.   Note that the vector unparticle contribution will become attractive for $\du < 1/2$, but this also results in a potential which weakens as $r \rightarrow 0$.  This is thus unlikely to greatly affect black hole formation.  It is nevertheless enlightening to examine possible black hole-like solutions that result from the modified potential, as the modified potential has a very important consequence: chargeless, spinless black holes can possess two horizons.

In particular, for the case $\du = 1.5$:
\beq
\Phi(r) = \frac{r_s}{r} \left(1-\left(\frac{R_{*v}}{4\pi^3 r}\right)\right)~,
\eeq
The corresponding horizon condition that results from insertion of this potential into the metric yields a Reissner-N\"ordstrom-type solution with two horizons
\beq
r_{H\pm} = \frac{r_s}{2} \left( 1 \pm \sqrt{1-\frac{R_{*-}}{\pi^3 r_s}}\;\right)~~,
\eeq
with the constraint $M_{BH} > R_{*-}/2\pi^3$.  It is thus possible in principle to distinguish such an object from its Schwarzschild equivalent through its temperature, which for a Riessner-N\"ordstrom black hole is \cite{rstempref}

\beq
T_H = \frac{r_+ - r_-}{4\pi r_+^2} = T_+ (1- \delta)
\label{rstemp}
\eeq
where $T_+$ is the Hawking temperature of a black hole with radius $r_+$, and
\beq
\delta  = \frac{r_-}{r_+} = \frac{1-\sqrt{1-\frac{R_{*-}}{\pi^3 r_s}}}{1+\sqrt{1-\frac{R_{*-}}{\pi^3 r_s}}}
\eeq

Although such temperatures are difficult to measure due to their extremely small magnitude, in principle such a deviation should be in some way detectable.  Orbits around the object might also be discernible from those of a regular black hole, but such a calculation is beyond the scope of this discussion.

\vskip .5cm
\noindent{\bf Acknowledgments}\\
The author thanks Justin Khoury and Robert Mann for insightful comments.  Gratitude is also extended to the Perimeter Institute for Theoretical Physics for their generous hospitality, at which a portion of this research was completed.  The author is supported by a Cottrell College Science Award from Research Corporation.

\end{document}